\documentclass[conference,10pt]{IEEEtran}

\usepackage{amsmath,amsfonts,amssymb}
\usepackage{graphicx}
\usepackage{cite}
\usepackage[hidelinks]{hyperref}
\usepackage{array}
\usepackage{booktabs}
\usepackage{multirow}
\usepackage{float}
\usepackage{caption}
\usepackage{subfig}
\usepackage{xcolor}

\graphicspath{{Figures/}}

\IEEEoverridecommandlockouts

\begin{document}

\title{Optimal FALQON for Quantum Approximate Optimization via Layer-wise Parameter Tuning}

\author{\IEEEauthorblockN{Michael Mancini and Shabnam Sodagari, \textit{Senior Member, IEEE}}
\IEEEauthorblockA{Computer Engineering \& Computer Science\\
California State University, Long Beach, CA 90840\\
michael@mancini.institute; shabnam@csulb.edu}}

\maketitle

\begin{abstract}
Feedback-based adaptive quantum optimization (FALQON) is a promising approach for combinatorial optimization on noisy intermediate-scale quantum devices, but prior studies report that standard fixed-parameter schedules can require relatively deep circuits to reach strong solution quality. We study Optimal FALQON, a layer-wise formulation that treats the time-step $\delta_k$ and scaling parameter $w_k$ as classical decision variables optimized per layer. The benchmark uses all 94 non-isomorphic 3-regular graphs with 12 vertices under 8192-shot simulation and compares standard FALQON, Optimal FALQON, and QAOA/QAOA-MA Powell runs warm-started from FALQON-derived schedules. Across approximation ratio, success probability, evaluation-normalized efficiency, and layer-normalized efficiency, Optimal FALQON substantially improves on standard FALQON, and Optimal-FALQON warm starts substantially improve on standard-FALQON warm starts.
\end{abstract}


\section{Introduction}
\label{sec:introduction}

Quantum optimization algorithms such as the Variational Quantum Eigensolver (VQE), the Quantum Approximate Optimization Algorithm (QAOA), and Feedback-Based Adaptive Quantum Optimization (FALQON) are actively studied for combinatorial optimization on noisy intermediate-scale quantum devices \cite{Peruzzo2014,McClean2016,Farhi2014,Magann2022,Arai2025}. QAOA relies on classical optimization to search for low-energy circuit parameterizations, often requiring multiple measured circuit executions. FALQON avoids a full outer-loop variational search, but prior studies report that standard FALQON can require relatively deep circuits when fixed control schedules are used \cite{Magann2022,Arai2025}. This is particularly problematic on quantum hardware where circuit depth and shot budgets are constrained.

Representative superconducting-qubit systems have historically exhibited non-negligible single- and two-qubit gate errors, motivating depth-conscious algorithm design \cite{Arute2019,Kjaergaard2020}. Recent FALQON variants therefore modify the original fixed schedule through time rescaling, robust gain regularization, and other feedback-control accelerations \cite{Rattighieri2025,Legnini2025}. These results motivate directly optimizing the layer-wise control schedule itself.

This paper studies Optimal FALQON, a layer-wise variant in which the time-step $\delta_k$ and scaling parameter $w_k$ are optimized per layer with Powell's method. We evaluate it on MaxCut over all 94 non-isomorphic 3-regular graphs with 12 vertices and then test whether the resulting FALQON schedules are useful warm starts for QAOA and multi-angle QAOA under Powell optimization.

The main contributions are:
\begin{itemize}
	\item A layer-wise Optimal FALQON procedure that uses bounded Powell search to tune the cost-evolution timestep $\delta_k$ and feedback gain $w_k$ for fixed first- and second-order feedback laws.
	
	\item A complete $N=12$ 3-regular MaxCut benchmark showing that Optimal FALQON improves approximation ratio, success probability, and cost-aware efficiency over fixed-schedule FALQON.
	
	\item A warm-start study showing that Optimal-FALQON schedules improve QAOA and QAOA-MA initialization over standard-FALQON schedules, with matched-instance Wilcoxon-Holm tests supporting the principal improvements.
\end{itemize}

\section{Related Work and Background}
\label{sec:background}

VQE established the variational hybrid quantum-classical paradigm, and QAOA adapted that framework to combinatorial optimization with a layer-depth-efficient alternating ansatz \cite{Peruzzo2014,McClean2016,Farhi2014}. QAOA-MA extends standard QAOA by assigning separate angles to individual Hamiltonian terms, increasing expressiveness at the cost of a larger classical search space \cite{Herrman2022}.

FALQON introduced a feedback-based alternative that uses commutator measurements to generate layer-wise control updates without a full outer-loop variational search \cite{Magann2022,MagannPRA2022}. Subsequent work introduced time-rescaled schedules, robust gain regularization, quadratic approximations for depth reduction, and related control accelerations such as FOCQS and GD-QLC \cite{Rattighieri2025,Legnini2025,Arai2025,Brady2024,Mozakka2026}. Magann et al. also discussed seeding QAOA from FALQON-generated parameters \cite{MagannPRA2022}; here we test whether seeding from Optimal FALQON improves that warm-start strategy.

Quantum circuits operate under shot-noise limitations, introducing stochastic variation in measurement outcomes. PennyLane and related frameworks enable simulation and optimization of quantum circuits while preserving stochastic shot noise \cite{bergholm2018pennylane}.

\subsection{MaxCut Benchmark}

The MaxCut problem is a standard benchmark for variational quantum algorithms \cite{Farhi2014,Herrman2022,MuozArias2024}. For an undirected graph $G=(V,E)$ with $|V|=N$, we use the Ising-form problem Hamiltonian
\begin{equation}
H_p = \frac{1}{2}\sum_{(i,j) \in E}\left(Z_i Z_j - I\right).
\end{equation}
Minimizing $\langle H_p \rangle$ corresponds to maximizing the cut value. The benchmark set is the full collection of 94 non-isomorphic 3-regular graphs with $N=12$ vertices \cite{sagemath,mckay1981practical}. Exact optimal solutions are available by exhaustive enumeration over all $2^{12}=4096$ bitstrings.

\section{Optimal FALQON Method}
\label{sec:method}

\subsection{Layer-Wise Formulation}

FALQON uses commutator expectations
\begin{align}
A_k &= \langle \psi_k | i[H_d,H_p] | \psi_k \rangle, \\
B_k &= \left\langle \psi_k \left| \frac{1}{2}[[H_d,H_p],H_d] \right| \psi_k \right\rangle, \\
C_k &= \langle \psi_k | [[H_d,H_p],H_p] | \psi_k \rangle,
\end{align}
where $H_d$ is the driver Hamiltonian. Standard FALQON applies fixed global hyperparameters. In Optimal FALQON, the layer-$k$ control variables are instead optimized as
\begin{equation}
\delta_k \in \mathbb{R}, \quad w_k \in \mathbb{R}, \quad k \in [1,L].
\end{equation}

Starting from the uniform superposition state
\begin{equation}
|\psi_0\rangle = |+\rangle^{\otimes N},
\end{equation}
each layer applies the cost and mixer evolutions
\begin{equation}
|\psi_k\rangle = e^{-i\beta_k^{\mathrm{ctrl}}\delta_k H_d} e^{-i\delta_k H_p} |\psi_{k-1}\rangle.
\end{equation}
Thus $\delta_k$ is the cost-evolution step and $\beta_k^{\mathrm{ctrl}}\delta_k$ is the mixer-evolution angle actually applied in the circuit.

For the first-order rule we use
\begin{equation}
\beta_k^{\mathrm{ctrl,FO}} = -w_k A_{k-1}.
\end{equation}
For the second-order rule we form
\begin{equation}
\beta_k^{\mathrm{ctrl,SO}} =
\begin{cases}
-w_k\left(A_{k-1}+C_{k-1}\delta_k\right), & B_{k-1}=0, \\
-w_k\dfrac{A_{k-1}+C_{k-1}\delta_k}{2B_{k-1}\delta_k}, & B_{k-1}>0, \\
\phantom{-}w_k\dfrac{A_{k-1}+C_{k-1}\delta_k}{2B_{k-1}\delta_k}, & B_{k-1}<0.
\end{cases}
\end{equation}
Following the quadratic-approximation implementation of Arai et al., the second-order update is capped by the first-order update whenever the second-order step is larger in magnitude, i.e.,
\begin{equation}
\beta_k^{\mathrm{ctrl}} =
\begin{cases}
\beta_k^{\mathrm{ctrl,SO}}, & |\beta_k^{\mathrm{ctrl,SO}}| < |\beta_k^{\mathrm{ctrl,FO}}|, \\
\beta_k^{\mathrm{ctrl,FO}}, & \text{otherwise}.
\end{cases}
\end{equation}

At each layer, the classical optimization problem is
\begin{equation}
(\delta_k^{\ast}, w_k^{\ast}) = \arg\min_{\delta_k,w_k}\langle \psi_k(\delta_k,w_k)|H_p|\psi_k(\delta_k,w_k)\rangle.
\end{equation}
The Powell search is initialized at $x_k^{(0)}=(w_k,\delta_k)=(1,0.5)$ for every optimized layer in the present study.

Powell's method is a derivative-free optimizer \cite{powell1964efficient}. In the noisy VQLS optimizer benchmark of Pellow-Jarman et al., Powell was among the strongest optimizers, while Nelder-Mead and COBYLA were more noise-sensitive \cite{pellowjarman2021comparison}. That result does not prove universal optimality, but it motivates our use of Powell here.

\subsection{Benchmark Design}

The benchmark is performed on the MaxCut problem over the 94 non-isomorphic 3-regular graphs with $N=12$ vertices, using depths $L \in \{1,\ldots,10\}$ and 8192 shots per circuit evaluation. The fixed FALQON baselines use the conservative schedules $\delta=0.03, w=1$ for FO and $\delta=0.05, w=1$ for SO. For Optimal FALQON, each layer-wise Powell search was initialized at $(w_k,\delta_k)=(1,0.5)$ and limited to 20 objective-function evaluations per layer. This configuration was chosen for reproducibility of the benchmark and is not claimed to be optimal or appropriate beyond the benchmark setting.

For warm-started QAOA, the seed at depth $L$ is taken directly from the FALQON schedule,
\begin{equation}
\gamma_k^{(0)} = \delta_k, \qquad \beta_k^{(0)} = \beta_k^{\mathrm{ctrl}}\delta_k, \qquad k=1,\ldots,L.
\end{equation}
For warm-started QAOA-MA, the same layer values are repeated across all cost-term and mixer-term angles at that layer. After this initialization, QAOA and QAOA-MA are optimized with Powell using the same 20-evaluation search budget.

\begin{table}[h]
\centering
\caption{Benchmark configurations used in the reported study.}
\label{tab:benchmark_setup}
\scriptsize
\setlength{\tabcolsep}{3pt}
\begin{tabular}{p{0.18\columnwidth}p{0.29\columnwidth}p{0.16\columnwidth}p{0.14\columnwidth}c}
\toprule
Method & Control Schedule & Search Budget & Classical Search & Shots \\
\midrule
FALQON FO & fixed $\delta=0.03$, $w=1$ & none & none & 8192 \\
FALQON SO & fixed $\delta=0.05$, $w=1$ & none & none & 8192 \\
Optimal FALQON FO & init $\delta_k=0.5$, $w_k=1$ & 20 evals/layer & Powell & 8192 \\
Optimal FALQON SO & init $\delta_k=0.5$, $w_k=1$ & 20 evals/layer & Powell & 8192 \\
Warm-start QAOA & seeded from FALQON & 20 evals & Powell & 8192 \\
Warm-start QAOA-MA & seeded from FALQON & 20 evals & Powell & 8192 \\
\bottomrule
\end{tabular}
\normalsize
\end{table}

\subsection{Performance Metrics}

We report four metrics. The approximation ratio is
\begin{equation}
r_A = \frac{\langle H_p \rangle}{(H_p)_{\min}},
\end{equation}
where $(H_p)_{\min}$ is the exact ground-state energy for the instance.

The probability that a single shot samples an optimal MaxCut solution is
\begin{equation}
P_{\text{success}} = \sum_{x^{(i)} \in \mathcal{S}^{\star}} P(x^{(i)}).
\end{equation}
Here $\mathcal{S}^{\star}$ is the set of optimal bitstrings for a graph.

We track the total number of circuit evaluations, $n_{\text{evals}}$, and use it to define the reported efficiency metrics:
\begin{equation}
E_1 = \frac{P_{\text{success}}}{n_{\text{evals}}},
\end{equation}
\begin{equation}
E_2 = \frac{P_{\text{success}}}{n_{\text{evals}} \times L}.
\end{equation}
Thus $E_1$ is evaluation-normalized success probability, and $E_2$ further normalizes by algorithmic layer count. In this study, $L$ is used as a layer-level proxy rather than a compiled hardware gate-depth metric.

\subsection{Statistical Methodology}

For each depth and metric, we compare matched graph-instance outcomes with paired Wilcoxon signed-rank tests \cite{wilcoxon1945,hollander2013nonparametric}. The Wilcoxon signed-rank test is a non-parametric paired test of whether the median of the pairwise differences is zero. Holm correction is then applied within each metric family to control family-wise error at $\alpha=0.05$ \cite{holm1979}. We report depth-wise significance counts together with the overall median values so that the statistical claims remain tied to the per-layer tests actually performed.

\section{Simulation Results}
\label{sec:results}

\subsection{FALQON Family}

\begin{figure*}[t]
\centering
\subfloat[$r_A$ for the FALQON family.]{\includegraphics[width=0.49\textwidth]{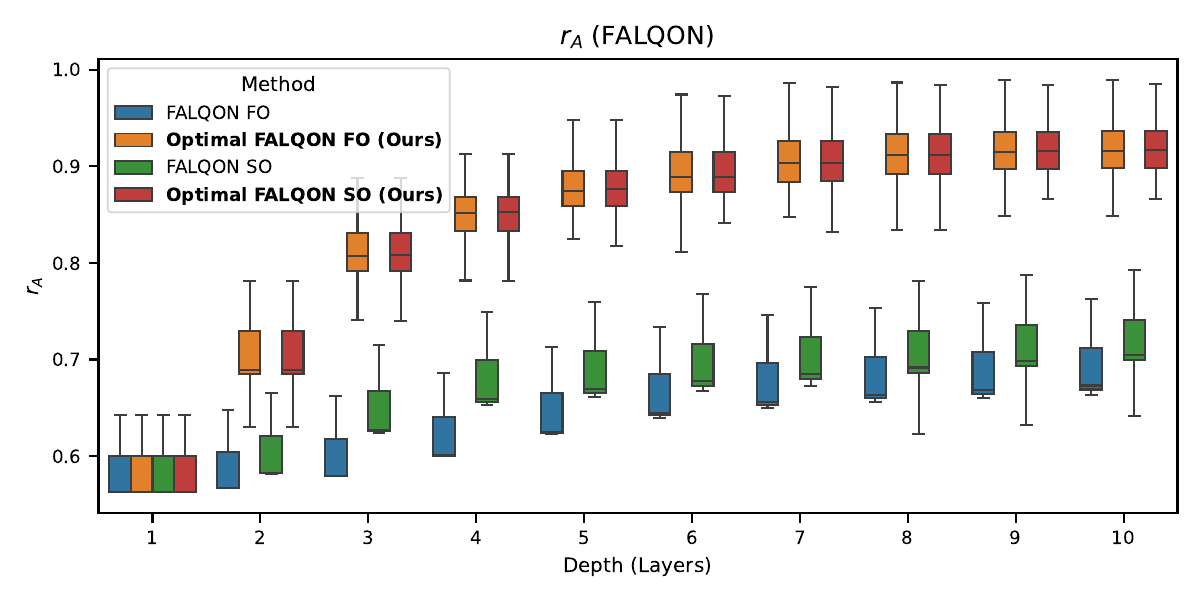}\label{fig:falqon_approx_ratio}}
\hfill
\subfloat[$P_{\text{success}}$ for the FALQON family.]{\includegraphics[width=0.49\textwidth]{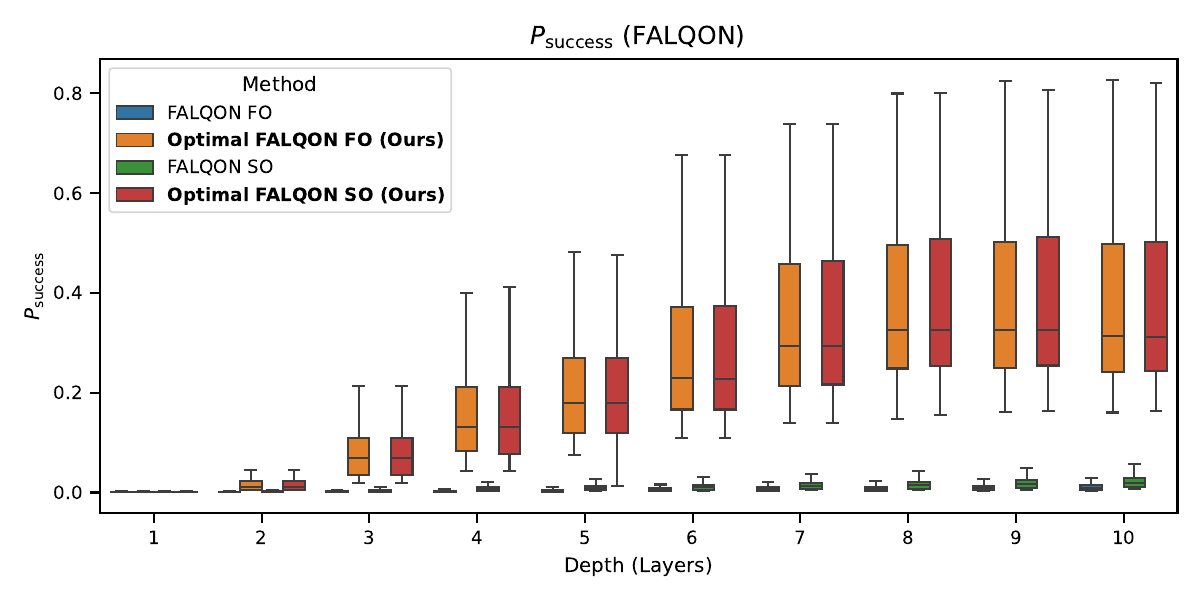}\label{fig:falqon_success}}
\caption{Depth-wise distributions for the FALQON family. Layer-wise schedule optimization improves both approximation ratio and success probability over fixed FALQON.}
\label{fig:falqon_family}
\end{figure*}

Figure~\ref{fig:falqon_family} shows the central within-family result. After depth $L=2$, the optimized schedules separate clearly from the fixed baselines in both approximation ratio and success probability. At depth $L=10$, the median success probability reaches $3.09\times 10^{-1}$ for both Optimal FALQON variants, compared with $9.78\times 10^{-3}$ for FALQON FO and $1.95\times 10^{-2}$ for FALQON SO.

The aggregate medians in Table~\ref{tab:overall_summary} show the same pattern. Standard FALQON yields overall median approximation ratios of $0.6347$ (FO) and $0.6734$ (SO), whereas Optimal FALQON reaches $0.8820$ (FO) and $0.8826$ (SO). The overall median success probability increases from $4.56\times 10^{-3}$ and $9.55\times 10^{-3}$ to $2.04\times 10^{-1}$ and $2.03\times 10^{-1}$, respectively. The efficiency metrics also improve strongly: the median $E_1$ rises from about $2\times 10^{-4}$ to at least $1.25\times 10^{-3}$, and the median $E_2$ rises from about $3.6\times 10^{-5}$ to at least $1.99\times 10^{-4}$.

\subsection{Warm-Started QAOA and QAOA-MA (Powell)}

\begin{figure*}[t]
\centering
\subfloat[$r_A$ for warm-started QAOA with Powell optimization.]{\includegraphics[width=0.49\textwidth]{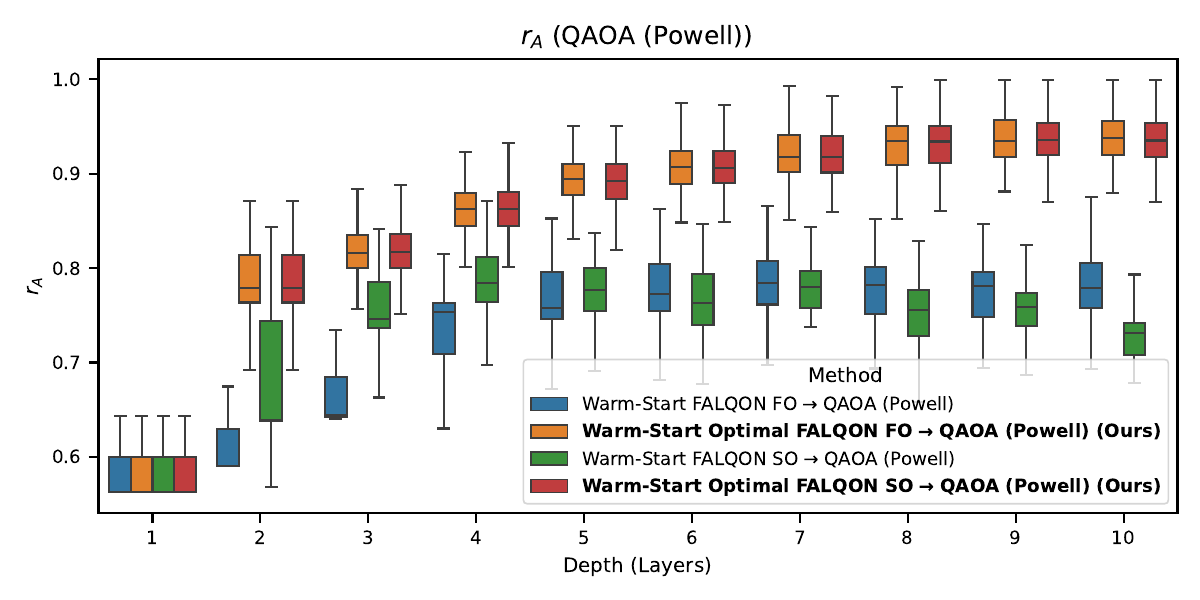}\label{fig:qaoa_approx_ratio}}
\hfill
\subfloat[$P_{\text{success}}$ for warm-started QAOA with Powell optimization.]{\includegraphics[width=0.49\textwidth]{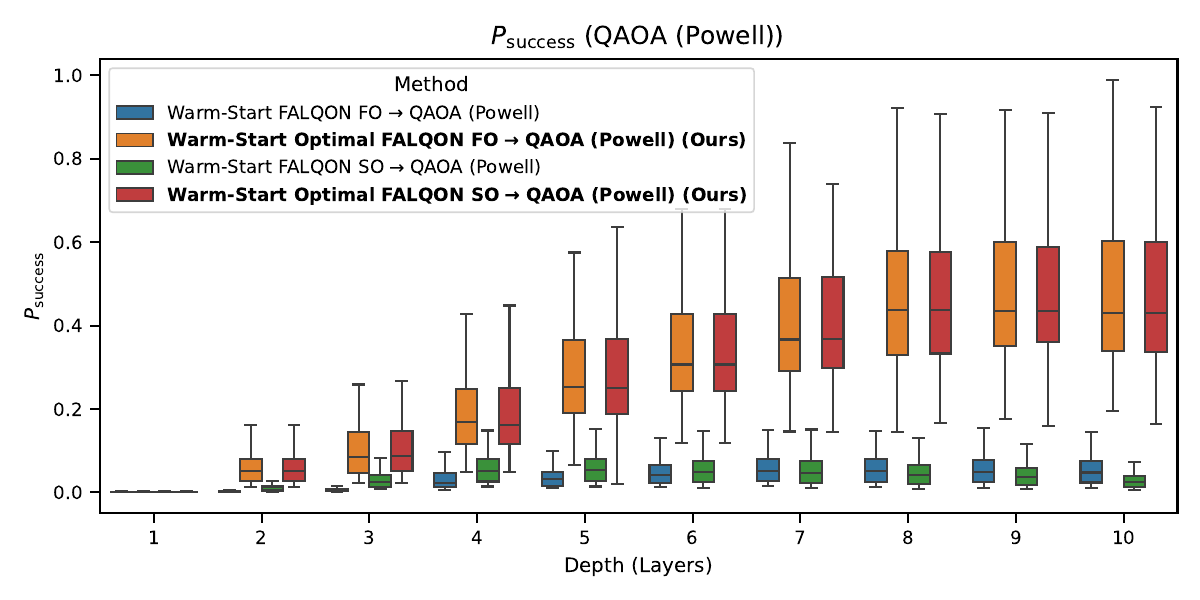}\label{fig:qaoa_success}}
\caption{Depth-wise distributions for warm-started QAOA with Powell optimization. Optimal-FALQON warm starts dominate standard-FALQON warm starts on both solution quality and success probability.}
\label{fig:qaoa_family}
\end{figure*}

Figure~\ref{fig:qaoa_family} summarizes warm-started QAOA under Powell optimization. Replacing a standard-FALQON warm start with an Optimal-FALQON warm start shifts the distributions upward across the tested depths in both approximation ratio and success probability. The overall median success probability rises from $3.71\times 10^{-2}$ and $3.96\times 10^{-2}$ for standard-FALQON warm starts to $2.80\times 10^{-1}$ and $2.78\times 10^{-1}$ for the Optimal-FALQON warm starts.

\begin{figure*}[t]
\centering
\subfloat[$r_A$ for warm-started QAOA-MA with Powell optimization.]{\includegraphics[width=0.49\textwidth]{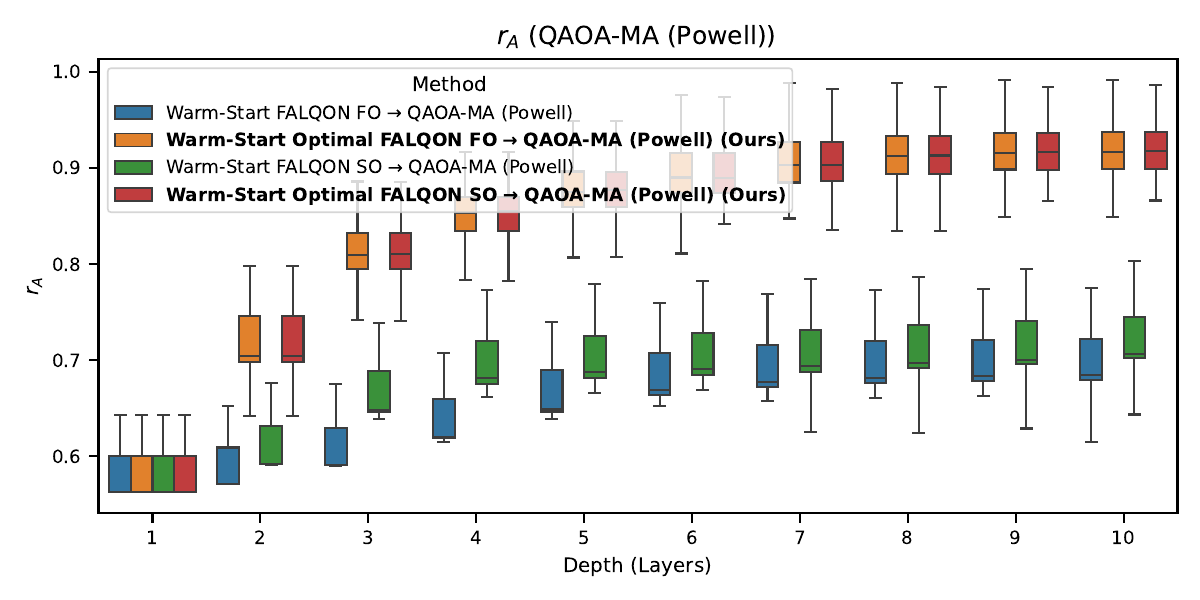}\label{fig:qaoa_ma_approx_ratio}}
\hfill
\subfloat[$P_{\text{success}}$ for warm-started QAOA-MA with Powell optimization.]{\includegraphics[width=0.49\textwidth]{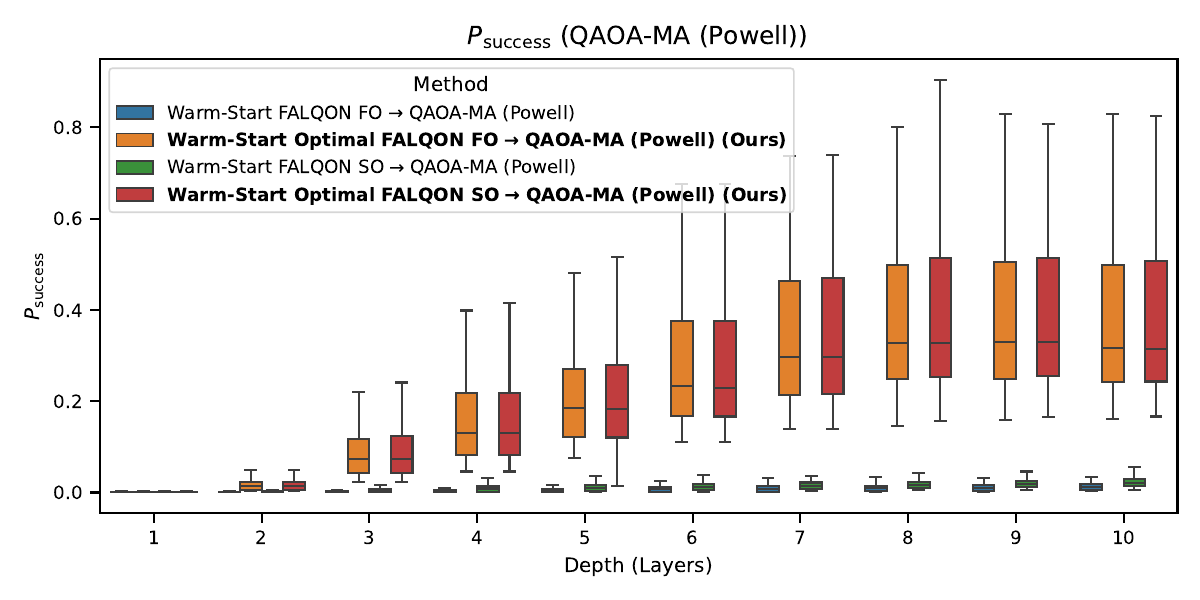}\label{fig:qaoa_ma_success}}
\caption{Depth-wise distributions for warm-started QAOA-MA with Powell optimization. Optimal-FALQON warm starts again improve both approximation ratio and success probability.}
\label{fig:qaoa_ma_family}
\end{figure*}

Figure~\ref{fig:qaoa_ma_family} shows the corresponding QAOA-MA benchmark. The same warm-start effect remains visible, but single-angle QAOA warm-started from Optimal FALQON is stronger in this particular dataset than warm-started QAOA-MA. The overall median success probability rises from $5.68\times 10^{-3}$ and $1.17\times 10^{-2}$ for standard-FALQON warm starts to $2.09\times 10^{-1}$ and $2.06\times 10^{-1}$ for the Optimal-FALQON warm starts.

The approximation-ratio results follow the same trend. Warm-started QAOA reaches overall median approximation ratios of $0.9009$ and $0.8992$ when seeded from Optimal FALQON, while the corresponding standard-FALQON warm starts remain near $0.757$ and $0.765$. Warm-started QAOA-MA reaches overall median approximation ratios near $0.8825$ and $0.8829$, substantially above the standard-FALQON warm-start baselines near $0.659$ and $0.689$.

\begin{figure*}[t]
\centering
\subfloat[FALQON: $E_1$.]{\includegraphics[width=0.49\textwidth]{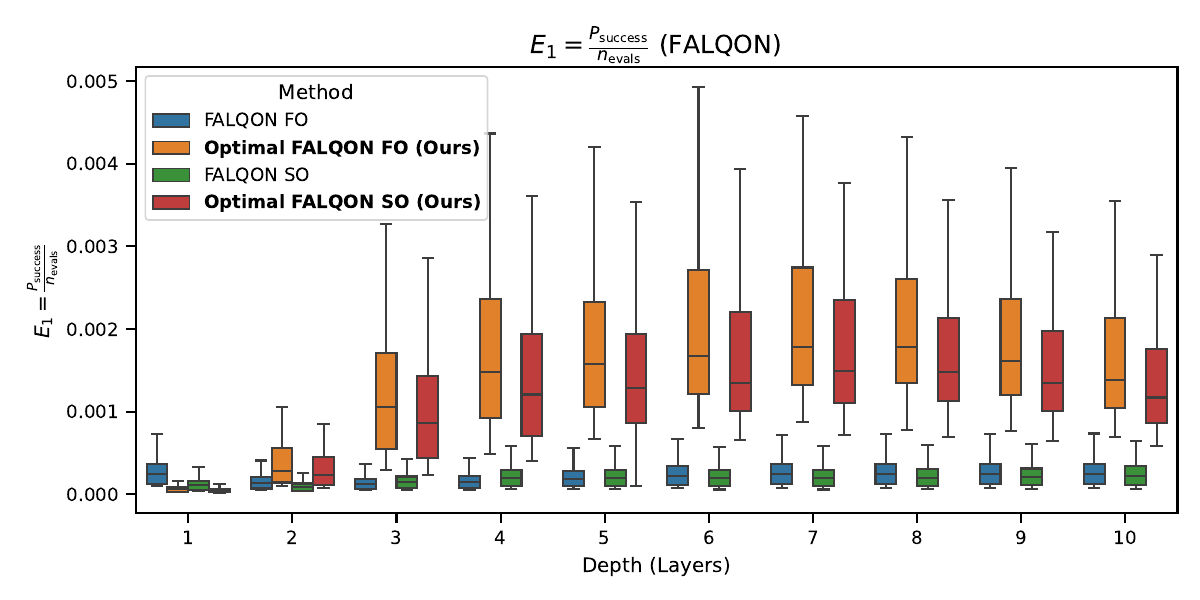}\label{fig:falqon_e1}}
\hfill
\subfloat[FALQON: $E_2$.]{\includegraphics[width=0.49\textwidth]{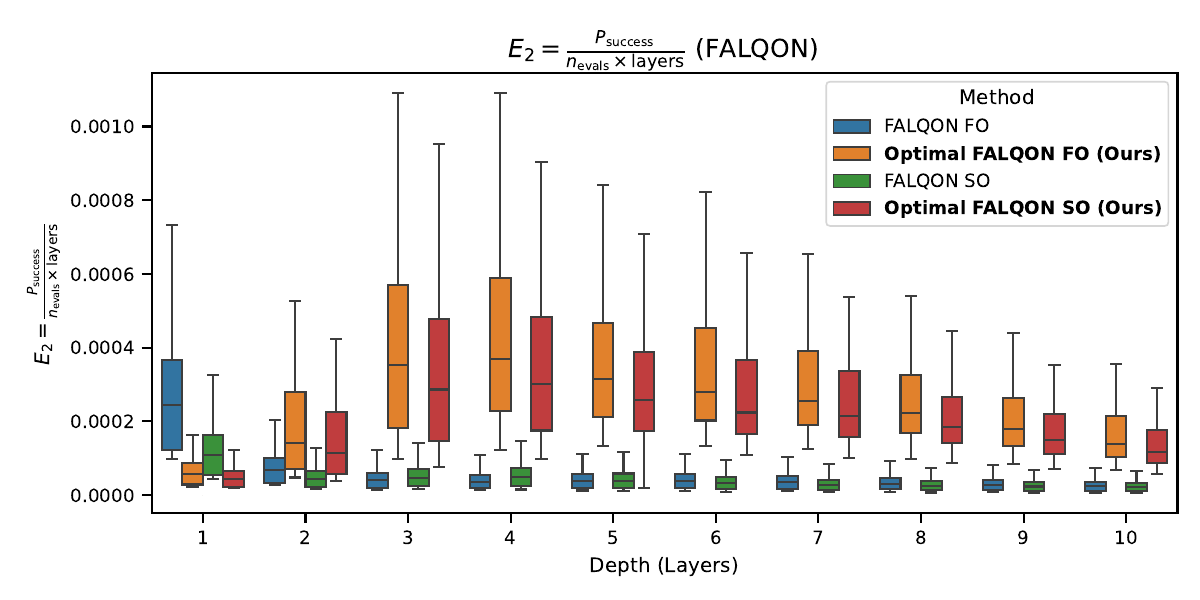}\label{fig:falqon_e2}}\\
\subfloat[QAOA (Powell): $E_1$.]{\includegraphics[width=0.49\textwidth]{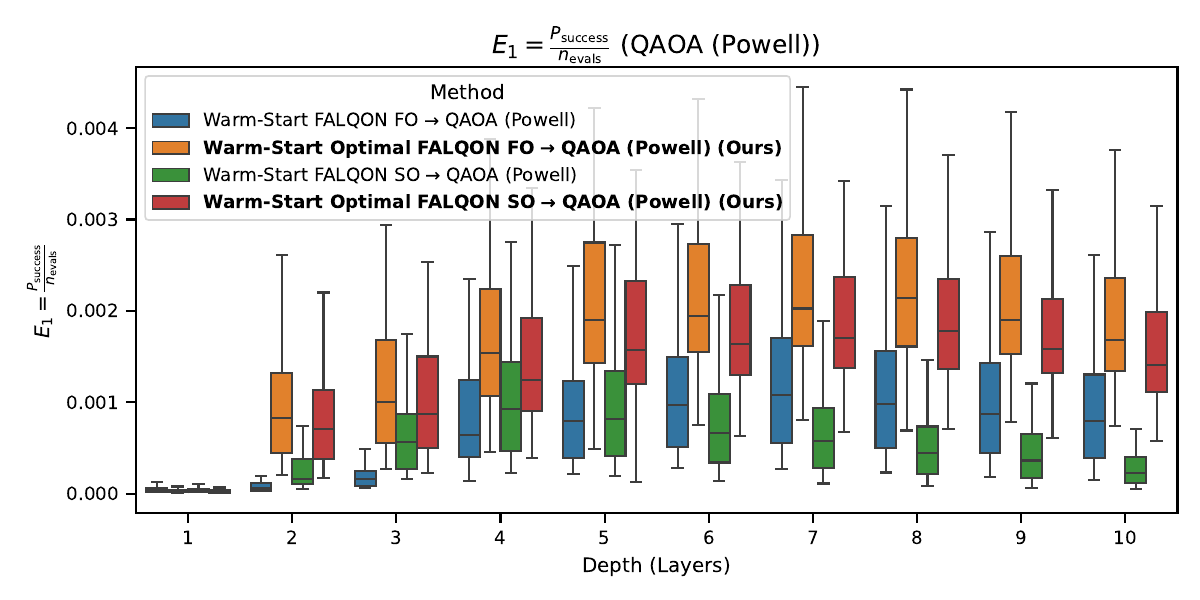}\label{fig:qaoa_e1}}
\hfill
\subfloat[QAOA (Powell): $E_2$.]{\includegraphics[width=0.49\textwidth]{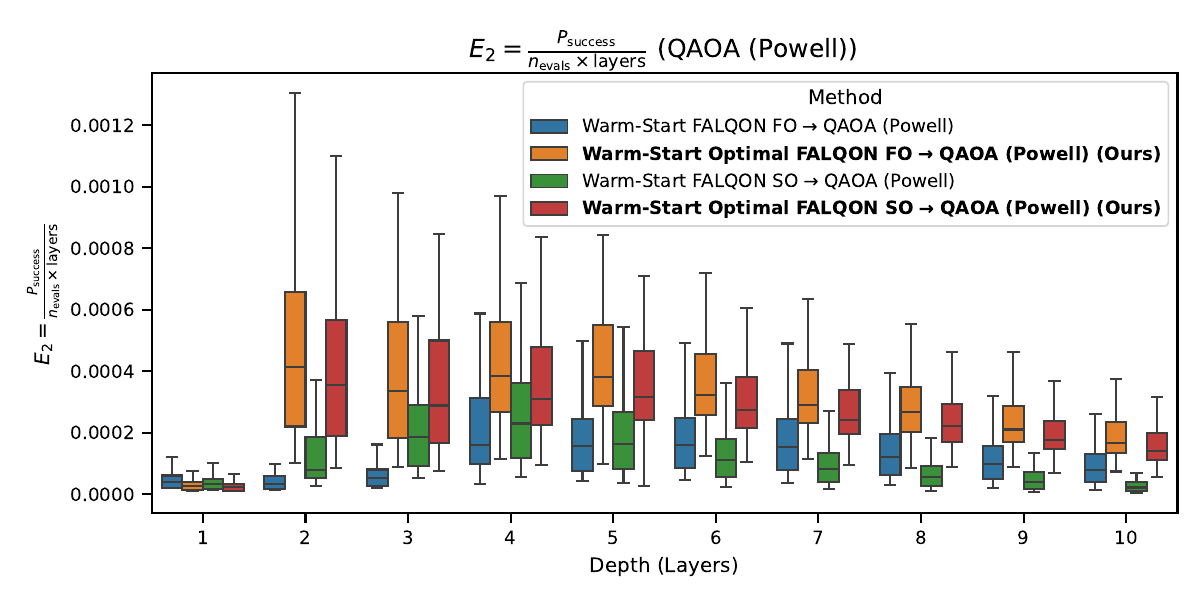}\label{fig:qaoa_e2}}\\
\subfloat[QAOA-MA (Powell): $E_1$.]{\includegraphics[width=0.49\textwidth]{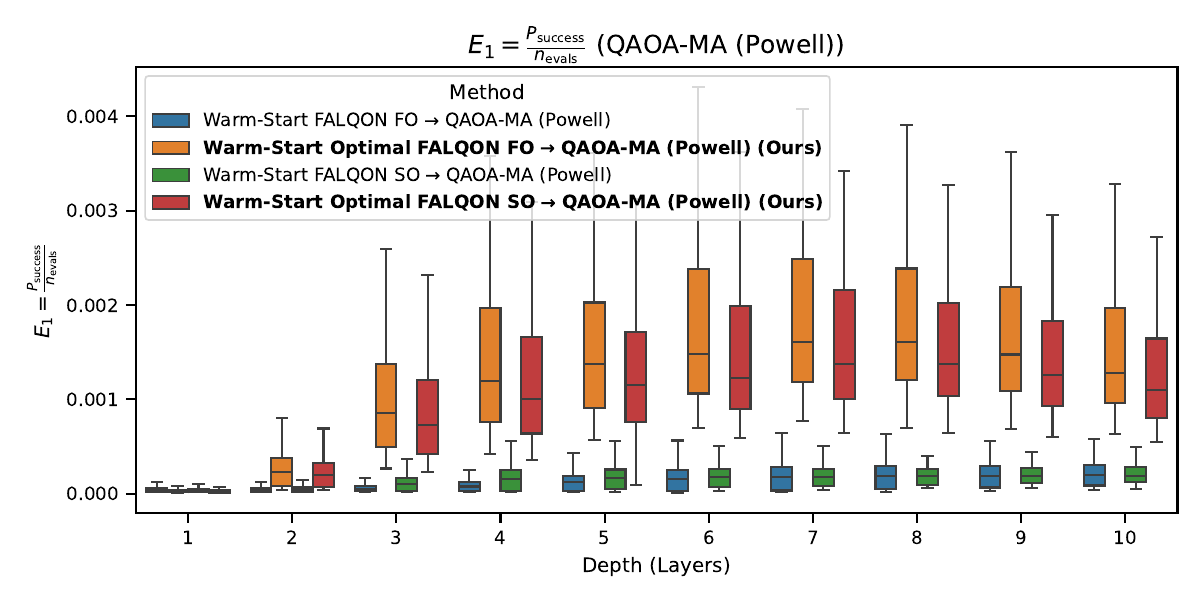}\label{fig:qaoa_ma_e1}}
\hfill
\subfloat[QAOA-MA (Powell): $E_2$.]{\includegraphics[width=0.49\textwidth]{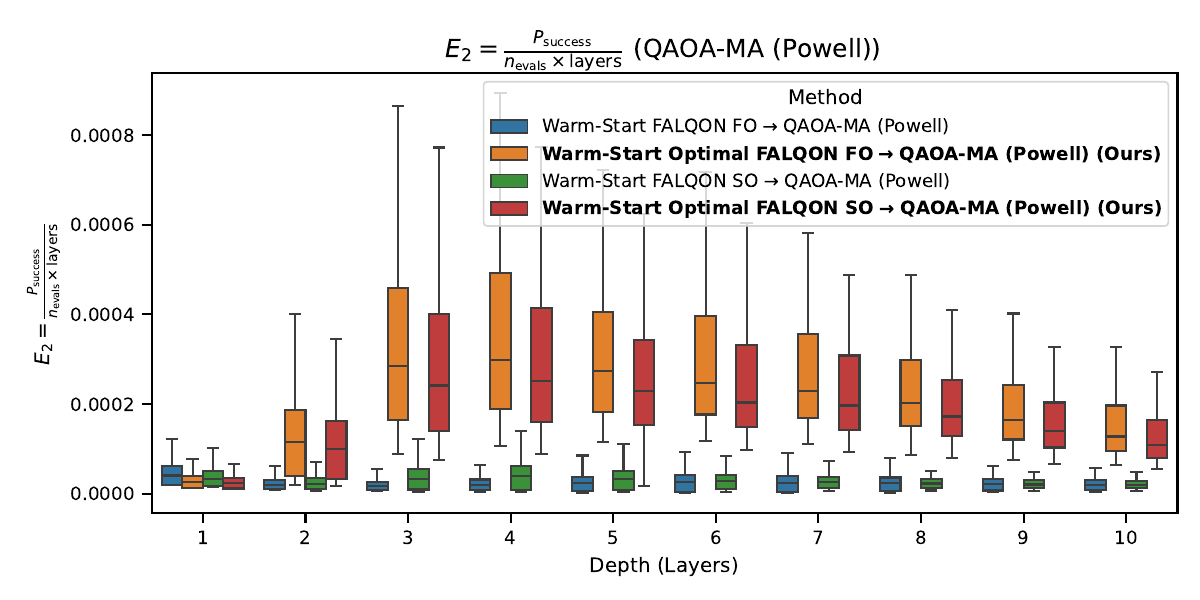}\label{fig:qaoa_ma_e2}}
\caption{Depth-wise efficiency distributions for the three benchmark families. The $E_1$ panels report evaluation-normalized success probability, and the $E_2$ panels additionally normalize by layer count.}
\label{fig:efficiency_family}
\end{figure*}

Figure~\ref{fig:efficiency_family} complements the solution-quality view by showing all six efficiency plots. Within the FALQON family, Optimal FALQON improves strongly over fixed FALQON in both $E_1$ and $E_2$ after the shallowest layer. Within the warm-started QAOA and QAOA-MA families, the same overall pattern remains visible: Optimal-FALQON warm starts improve the efficiency metrics at most depths, while the shallowest layer can still favor the corresponding standard-FALQON warm start.

\begin{table*}[t]
\centering
\caption{Overall median performance for the reported study.}
\label{tab:overall_summary}
\scriptsize
\setlength{\tabcolsep}{3pt}
\begin{tabular}{lcccc}
\toprule
Method & Median $r_A$ & Median $P_{\text{success}}$ & Median $E_1$ & Median $E_2$ \\
\midrule
FALQON FO & $6.3472 \times 10^{-1}$ & $4.5626 \times 10^{-3}$ & $2.3338 \times 10^{-4}$ & $3.6751 \times 10^{-5}$ \\
FALQON SO & $6.7342 \times 10^{-1}$ & $9.5539 \times 10^{-3}$ & $1.9497 \times 10^{-4}$ & $3.5537 \times 10^{-5}$ \\
$\textbf{Optimal FALQON FO}$ & $8.8195 \times 10^{-1}$ & $2.0425 \times 10^{-1}$ & $1.5290 \times 10^{-3}$ & $2.3867 \times 10^{-4}$ \\
$\textbf{Optimal FALQON SO}$ & $8.8263 \times 10^{-1}$ & $2.0326 \times 10^{-1}$ & $1.2467 \times 10^{-3}$ & $1.9945 \times 10^{-4}$ \\
Warm-Start FALQON FO $\to$ QAOA (Powell) & $7.6517 \times 10^{-1}$ & $3.7083 \times 10^{-2}$ & $7.9274 \times 10^{-4}$ & $1.0969 \times 10^{-4}$ \\
Warm-Start FALQON SO $\to$ QAOA (Powell) & $7.5692 \times 10^{-1}$ & $3.9627 \times 10^{-2}$ & $5.0282 \times 10^{-4}$ & $8.0176 \times 10^{-5}$ \\
Warm-Start $\textbf{Optimal FALQON FO}$ $\to$ QAOA (Powell) & $9.0090 \times 10^{-1}$ & $2.7982 \times 10^{-1}$ & $1.7892 \times 10^{-3}$ & $3.0669 \times 10^{-4}$ \\
Warm-Start $\textbf{Optimal FALQON SO}$ $\to$ QAOA (Powell) & $8.9925 \times 10^{-1}$ & $2.7836 \times 10^{-1}$ & $1.4933 \times 10^{-3}$ & $2.5796 \times 10^{-4}$ \\
Warm-Start FALQON FO $\to$ QAOA-MA (Powell) & $6.5876 \times 10^{-1}$ & $5.6811 \times 10^{-3}$ & $1.3807 \times 10^{-4}$ & $2.2105 \times 10^{-5}$ \\
Warm-Start FALQON SO $\to$ QAOA-MA (Powell) & $6.8936 \times 10^{-1}$ & $1.1677 \times 10^{-2}$ & $1.6780 \times 10^{-4}$ & $2.6869 \times 10^{-5}$ \\
Warm-Start $\textbf{Optimal FALQON FO}$ $\to$ QAOA-MA (Powell) & $8.8247 \times 10^{-1}$ & $2.0865 \times 10^{-1}$ & $1.3245 \times 10^{-3}$ & $2.1531 \times 10^{-4}$ \\
Warm-Start $\textbf{Optimal FALQON SO}$ $\to$ QAOA-MA (Powell) & $8.8287 \times 10^{-1}$ & $2.0623 \times 10^{-1}$ & $1.1225 \times 10^{-3}$ & $1.8452 \times 10^{-4}$ \\
\bottomrule
\end{tabular}
\end{table*}

\subsection{Statistical Summary}

Table~\ref{tab:stats_summary} summarizes the layer-wise Wilcoxon-Holm outcomes for the principal pairwise comparisons, reporting how many of the 10 tested depths were significant and which method was favored on those significant layers.

Within the FALQON family, both Optimal FALQON variants dominate their corresponding standard baselines in approximation ratio and success probability on all tested layers. The efficiency metrics show the same overall result, although the shallowest layer favors the corresponding standard baseline while the remaining significant layers favor Optimal FALQON.

The warm-start comparisons show the same dominant trend. For both QAOA and QAOA-MA under Powell optimization, replacing a standard-FALQON warm start with an Optimal-FALQON warm start yields broad layer-wise gains in approximation ratio and success probability. The efficiency metrics again favor the Optimal-FALQON warm starts on most significant layers, with shallow-depth exceptions visible in several comparisons.

The FO-SO comparison within Optimal FALQON is more selective. Second-order Optimal FALQON is significant only on a subset of layers for approximation ratio and at only one layer for success probability, whereas first-order Optimal FALQON is favored by the efficiency metrics on all tested layers.

Taken together, these layer-wise statistical results support the same central interpretation as the distributional plots: the main performance gain comes from adaptive layer-wise schedule optimization itself, while the first-order variant provides the more efficient practical configuration in the present benchmark.

\begin{table*}[t]
\centering
\caption{Depth-wise Wilcoxon-Holm summary for the principal pairwise comparisons. Each entry reports the number of significant layers out of 10 and the method favored on those significant layers. Here OF and SF denote Optimal FALQON and standard FALQON, while FO and SO denote first-order and second-order variants.}
\label{tab:stats_summary}
\scriptsize
\setlength{\tabcolsep}{3pt}
\begin{tabular}{p{0.42\textwidth}p{0.13\textwidth}p{0.13\textwidth}p{0.13\textwidth}p{0.13\textwidth}}
\toprule
Comparison & $r_A$ & $P_{\text{success}}$ & $E_1$ & $E_2$ \\
\midrule
OF-FO vs SF-FO & 10 (OF-FO) & 10 (OF-FO) & 9 (OF-FO), 1 (SF-FO) & 9 (OF-FO), 1 (SF-FO) \\
OF-SO vs SF-SO & 10 (OF-SO) & 10 (OF-SO) & 9 (OF-SO), 1 (SF-SO) & 9 (OF-SO), 1 (SF-SO) \\
OF-FO vs OF-SO & 4 (OF-SO) & 1 (OF-SO) & 10 (OF-FO) & 10 (OF-FO) \\
Warm-Start OF-FO $\to$ QAOA vs Warm-Start SF-FO $\to$ QAOA & 9 (OF-FO) & 10 (OF-FO) & 9 (OF-FO), 1 (SF-FO) & 9 (OF-FO), 1 (SF-FO) \\
Warm-Start OF-SO $\to$ QAOA vs Warm-Start SF-SO $\to$ QAOA & 10 (OF-SO) & 9 (OF-SO), 1 (SF-SO) & 9 (OF-SO), 1 (SF-SO) & 9 (OF-SO), 1 (SF-SO) \\
Warm-Start OF-FO $\to$ QAOA-MA vs Warm-Start SF-FO $\to$ QAOA-MA & 10 (OF-FO) & 10 (OF-FO) & 9 (OF-FO), 1 (SF-FO) & 9 (OF-FO), 1 (SF-FO) \\
Warm-Start OF-SO $\to$ QAOA-MA vs Warm-Start SF-SO $\to$ QAOA-MA & 9 (OF-SO) & 10 (OF-SO) & 9 (OF-SO), 1 (SF-SO) & 9 (OF-SO), 1 (SF-SO) \\
\bottomrule
\end{tabular}
\end{table*}

\section{Discussion and Conclusion}
\label{sec:conclusion}

Directly optimizing the FALQON control schedule through layer-wise $(\delta_k,w_k)$ search yields much stronger approximation ratios and success probabilities than fixed schedules, and those gains transfer downstream when the resulting schedules are used to warm-start QAOA and QAOA-MA.

Two practical conclusions stand out. First, the main performance gain comes from adaptive layer-wise schedule optimization. Second, the first-order Optimal FALQON variant is the preferable practical method in this benchmark: it remains stronger in the efficiency metrics, while the second-order variant is only intermittently distinguishable in approximation ratio and rarely distinguishable in success probability. In the downstream warm-start tests, QAOA with Powell and Optimal-FALQON seeding is the strongest performer in this dataset, while QAOA-MA still benefits substantially from the same seeding strategy.

These results suggest that Optimal FALQON is useful both as a standalone feedback-based optimizer and as a schedule generator for variational warm starts in near-term quantum optimization workflows.

\bibliographystyle{IEEEtran}
\bibliography{references}

\end{document}